\documentclass[aps,prl,showpacs,twocolumn,preprintnumbers,amsmath,amsfonts,bm]{revtex4}
\usepackage[dvips]{graphicx}
\usepackage{dcolumn}

\begin{document}
\title{Self-organization of heterogeneous topology and symmetry breaking in networks with adaptive thresholds and rewiring}

\author{Thimo Rohlf}

\affiliation{Santa Fe Institute, 1399 Hyde Park Road, Santa Fe, NM 87501, U.S.A }
\date{\today}

\begin{abstract}
We study an evolutionary algorithm that locally adapts thresholds and wiring in Random
Threshold Networks, based on measurements of a dynamical order parameter. %We introduce
A control parameter $p$ determines the probability of threshold adaptations vs. link rewiring.
For any $p < 1$, we find spontaneous symmetry breaking into a new class of self-organized networks,
characterized by a much higher average connectivity $\bar{K}_{evo}$ than networks without threshold adaptation ($p =1$).
While $\bar{K}_{evo}$ and evolved out-degree distributions are independent from $p$ for $p <1$, in-degree distributions become broader
when $p \to 1$, approaching a power-law. In this limit, time scale separation between
threshold adaptions and rewiring also leads to strong correlations between thresholds and in-degree.
Finally, evidence is presented that networks converge to self-organized criticality for large $N$.

\end{abstract}

\maketitle

Interaction networks in nature often exhibit highly inhomogeneous architectures.
Examples are scale-free degree distributions in protein networks \cite{MaslovSneppen2002} and
metabolic networks \cite{Jeong2000}, mostly accompanied by intricate second order
regularities as, for example, community structure \cite{Girvan2002}. 
%or degree-degree correlations \cite{}.
The emergence of these properties often is explained by means of intuitive topology-based models, e.g.
preferential attachment \cite{Barabasi99} or node duplications \cite{Bebek2006}. Real networks,
however, are characterized not only by an evolving topology, but also by evolution of
{\em function}, conveniently abstracted in terms of dynamics, i.e. the flow of information or matter
on these networks. So far, only few studies explicitly consider the more general case of 
co-evolution between network dynamics and -topology \cite{BornholSneppen98,BornholRohlf00,BornholRoehl2003,LiuBassler2006}.

One example is the question how networks may evolve topologies that optimize
biologically relevant parameters, e.g. flexible adaptation with respect to
changing environments, or insensitivity against random perturbations of topology
or dynamics (robustness) \cite{Savageau71}. In this context, Kauffman 
introduced random Boolean networks (RBN)
to study the dynamics of gene regulatory networks from a global perspective \cite{Kauffman69,Kauffman93}.
It was shown that RBN undergo a order-disorder transition at a critical wiring density (connectivity) $K_c = 2$
\cite{Kauffman69,Kauffman93,DerridaP86,SoleLuque95}; similar results were established for random threshold networks (RTN), which constitute
a sub-class of RBN \cite{Kuerten88,RohlfBornhol02,Rohlf07}. It has been postulated that evolution should drive dynamical networks
towards this 'edge of chaos' to optimize adaptive flexibility and robustness \cite{Kauffman69,Kauffman93}. However,
%for a long time 
no mechanism able to generate critically connected networks could be provided.

To address this problem, a RTN-based model was proposed, linking rewiring of network nodes to local measurements
of a dynamical order parameter, e.g. the average activity (magnetization) \cite{BornholRohlf00}.
It was shown that this simple, {\em local} adaptive mechanism leads to a {\em global} self-organized critical state
in the limit of large system sizes $N$. Subsequently, this principle
was generalized to networks of noisy neurons \cite{BornholRoehl2003} and to RBN with evolvable logical
functions \cite{LiuBassler2006}. Interestingly, finite size networks in these models evolve a broadly
distributed heterogeneous in-degree connectivity \cite{LiuBassler2006,RohlfBornhol04}. Still, these topological heterogeneities
are smaller than those observed in real-world networks, presumably 
because dynamical elements were assumed to be homogeneous with respect to 
their dynamical behavior.
%activation thresholds (RTN)
%or randomly assigned Boolean functions (RBN). 
While this assumption leads to elegant 
models, it is quite unrealistic, as it becomes apparent e.g. in
the frequent occurrence of canalizing functions in gene regulatory networks, 
with strong impact on dynamics in RBN models \cite{Moreira05}. 
%Similarly, a recent study
%showed that dynamics of RTN is strongly affected by inhomogeneous thresholds \cite{Rohlf07}.
Considering the accumulating experimental evidence of both close-to criticality \cite{RamoeKesseliYli06} and heterogeneous
architecture \cite{Tong2004} in real gene regulatory networks, it is fascinating to speculate about a mechanism
that might explain both observations: evolution of {\em local} dynamical heterogeneity and
{\em global} homeostasis.
%Hence, more realistic models of network evolution have to consider
%the possibility that network elements may differ with respect to their dynamical properties, and that these
%differences evolve in time.

For this purpose, we introduce a minimal model linking regulation of activation thresholds and rewiring
of network nodes in RTN to local measurements of a dynamical order parameter. 
%Adaptation of thresholds
%opens up for the possibility of units that become heterogeneous with respect to their {\em dynamical} properties:
%nodes with high thresholds are inert and switch their state only for few input configurations (similar to the effect of
%canalizing functions in RBN), whereas nodes with low thresholds are more likely to switch. 
A new control parameter
$p \in [0,1]$ determines the probability of rewiring vs. threshold adaptations.
We show that the symmetry of the evolutionary attractor for $p = 1$ (no threshold adaptations, rewiring only) is broken spontaneously for
any $p < 1$. This new universality class of self-organized networks exhibits a much higher average connectivity
$\bar{K}_{evo}$, compared to $p =1$ networks, however, with a value $\bar{K}_{evo}$ that is {\em insensitive} to $p$.
In-degree distributions become very broad, approaching a flat power-law tail $\sim k_{in}^{-3/4}$ for $p \to 1$. Further, we
establish the emergence of strong correlations between in-degree and thresholds in this limit, while
for small $p$, correlations are weak. This indicates that an adaptive time-scale separation, with rare events of dynamical diversification
and frequent rewiring, can lead to emergence of highly inhomogeneous topologies, without
the need for network growth (as, for example, in preferential attachment models). Finally, we present evidence that networks
with $p < 1$ converge to a critical state for large $N$, however, with a finite size scaling significantly different
from the one found for the case $p =1$.

{\em Dynamics.} We consider a network of $N$ randomly interconnected binary 
elements  with states $\sigma_i=\pm 1$. For each site $i$, its state 
at time $t+1$ is a function of the inputs it receives from other 
elements at time $t$ (synchronous updates):
\begin{equation}
\sigma_i (t+1)=
\begin{cases} 
+1  & \text{if  }  f_i(t) > 0\\
-1 &\text{else}
\end{cases}
\end{equation}
with 
\begin{equation}
f_i(t) = \sum_{j=1}^N c_{ij}\sigma_j(t) + h_i.\label{statesum_eq}
\end{equation}
The interaction weights $c_{ij}$ 
take discrete values $c_{ij}=\pm1$, with $c_{ij} = 0$ if site 
$i$ does not receive any input from element $j$. Thresholds $h_i$
may vary from node to node, taking integer values $h_i \le 0$ \footnote{We chose $h_i \le 0$ to ensure that thresholds
make {\em activation}, i.e. $\sigma_i = +1$, more difficult.}. In the following discussion,
adaptive changes will be applied to the absolute value $|h_i|$, keeping in mind that
the sign of $h_i$ is always negative.

As a {\em dynamical order parameter}, we define the average activity $A(i)$ of a site $i$ 
\begin{eqnarray} 
A(i) = \frac{1}{T_2-T_1}\sum_{t=T_1}^{T_2}\sigma_i(t). 
\end{eqnarray}
Notice that a {\em frozen} site, i.e. a site that does not change its state, has $|A(i)| = 1$, whereas an {\em active} site
has $|A(i)| < 1$.

\begin{figure}[htb]
\begin{center}
\resizebox{65mm}{!}{\includegraphics{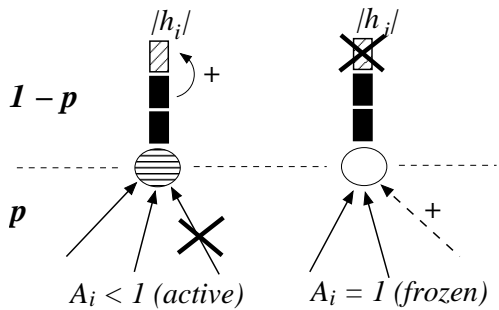}}
\end{center}
\caption{\small {\em Left:} with probability $p$, active nodes loose one of their inputs, with probability $1-p$ they increase their (absolute) threshold $|h_i|$.
{\em Right:} frozen nodes show the opposite behavior.  
%a node $i$ is selected at random and its activity $A_i$
%is determined. {\em Left panel:} If it is active ($A_i < 1$), with probability $p$ it looses one of its existing inputs, with probability
%$1-p$ the absolute value $|h_i|$ of its threshold is increased by one. {\em Right panel:} If the node is frozen ($A_i = 1$), with probability $p$ it receives an additional %input,
%with probability $1-p$ the absolute value $|h_i|$ of its threshold is decreased by one. 
}
\label{dofkfig} 
\end{figure}

\begin{figure}[htb]
\begin{center}
\resizebox{85mm}{!}{\includegraphics{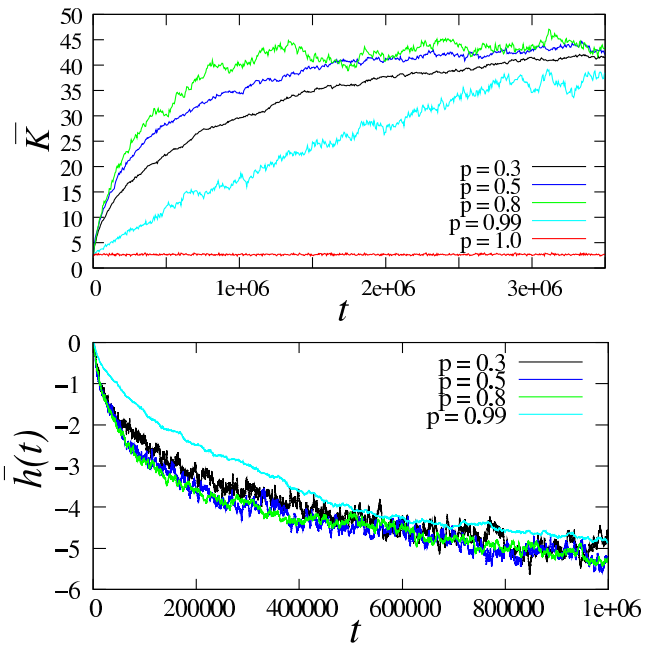}}
\end{center}
\caption{\small {\em Upper panel:} Evolution of the average connectivity $\bar{K}$ of threshold networks, using the adaptive algorithm 
(cf. Fig. 1), for $N = 512$ and initial connectivity $\bar{K}_{ini} = 1$. Time series for five different values of $p$ are shown. 
{\em Lower panel:} The same for the average threshold $\bar{h}$.}
\label{dofkfig} 
\end{figure}

\begin{figure}[htb]
\begin{center}
\resizebox{85mm}{!}{\includegraphics{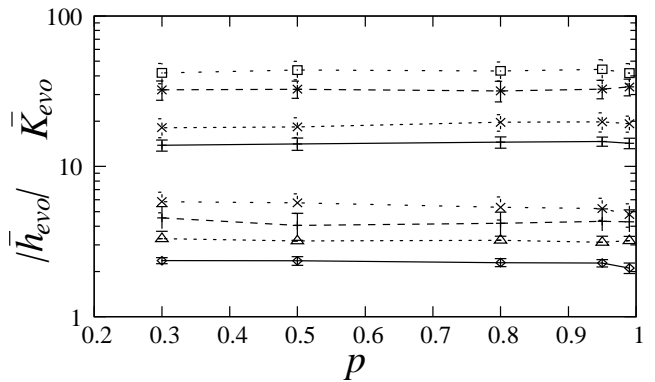}}
\end{center}
\caption{\small {\em Upper four curves:} Evolutionary mean values $\bar{K}_{evo}$ of the average connectivity, as a function of $p$; system sizes from top to bottom: $N=512$, $N=256$, $N=128$ and $N=64$. {\em Lower four curves:} The same for the
evolutionary mean values $|\bar{h}_{evo}| $of the average absolute threshold. Statistics was taken over $10^6$ evolutionary steps, after a transient of $4\cdot 10^6$ steps.}
\label{dofkfig} 
\end{figure}

\begin{figure}[htb]
\begin{center}
\resizebox{85mm}{!}{\includegraphics{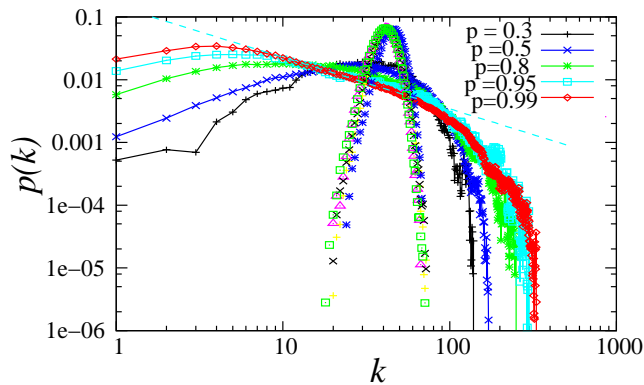}}
\end{center}
\caption{\small {\em Line-pointed curves:} in-degree distributions of evolved networks, {\em data points only:} the
corresponding out-degree distributions (($\triangle$) $p=0.3$, (+) $p =0.5$, (x) $p=0.8$, (*) $p =0.95$, ($\square$) $p=0.99$). 
Statistics was gathered over $10^6$ evolutionary steps, after
a transient of $4\cdot 10^6$ steps. Networks had size $N = 512$. The dashed line has slope $-3/4$. }
\label{dofkfig} 
\end{figure}

\begin{figure}[htb]
\begin{center}
\resizebox{85mm}{!}{\includegraphics{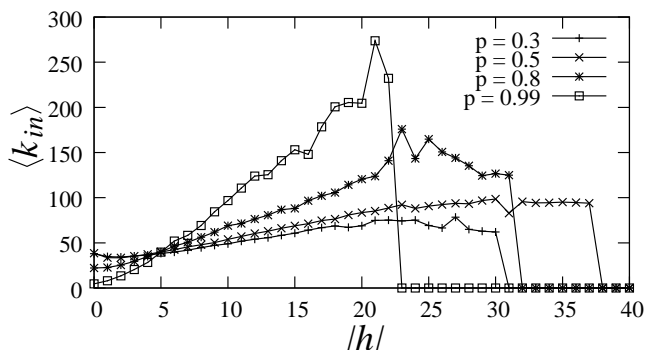}}
\end{center}
\caption{\small Average number $\langle k_{in}\rangle$ of inputs for a given node in evolving networks, as a function of the respective
nodes (absolute)
threshold $|h|$. Statistics was taken over $10^6$ rewiring steps, after a transient of $4\cdot 10^6$ steps. For all
values $p < 1$, a clear positive correlation between $\bar{k}_{in}$ and $|h|$ is found.}
\label{dofkfig} 
\end{figure}

\begin{figure}[htb]
\begin{center}
\resizebox{85mm}{!}{\includegraphics{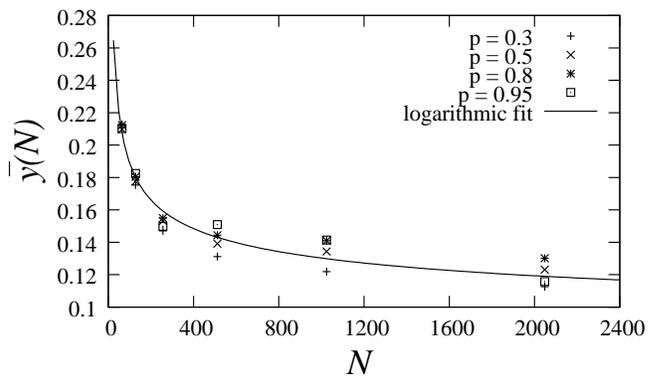}}
\end{center}
\caption{\small Average fraction $\bar{y}(N)$ of damaged nodes, 200 updates after a one-bit perturbation at time $t=0$, for different
$p$, as a function of system size $N$. The lined curve is a fit of the average scaling behavior.}
\label{dofkfig} 
\end{figure}

{\em Topology evolution.} Let us now discuss a particular evolutionary scheme
that couples local adaptations of both the number of inputs and of thresholds 
to a site's average activity.
Analyzing Eq. (1) and Eq. (\ref{statesum_eq}), we realize that the activity of a site $i$ 
can be controlled
in two ways: if $i$ is frozen, it can increase the probability to change its state
by either increasing its number of inputs $k_i \rightarrow k_i + 1$, or by making its threshold  $h_i \le 0$ less negative,
i.e. $|h_i| \rightarrow |h_i| - 1$. If $i$ is active, it can reduce its activity
by adapting either $k_i  \rightarrow k_i - 1$ or $|h_i| \rightarrow |h_i| + 1$. This adaptive scheme is realized in the following algorithm
(see also Fig. 1): \\
%\begin{enumerate}
1. Create a random network with average connectivity $\bar{K}_{ini} > 0$ and average threshold
$\bar{h}_{ini} = 0$.\\
2. Select a random initial state $\vec{\sigma}_{ini} = (\sigma_1,...,\sigma_N)$.\\
3. Iterate network dynamics for $T$ timesteps.\\
4. Select a network site $i$ at random and measure its average activity $A_i$ over the last
$T/2$ updates.\\
5. Adapt $k_i$ and $h_i$ in the following way:\\
- If $|A_i| < 1$, then $k_i \rightarrow k_i-1$ with probability $p$ (removal of one randomly selected input).
With probability $1-p$, adapt $|h_i| \rightarrow |h_i| + 1$ instead.\\   
- If $|A_i| = 1$, then $k_i \rightarrow k_i+1$ with probability $p$ (addition of a new input from a randomly selected site).
With probability $1-p$, adapt $|h_i| \rightarrow |h_i| - 1$ instead. If $h_i = 0$, let its value unchanged. \\
6. Go back to step 3.

%\end{enumerate}
If the control parameter $p$ takes values $p > 1/2$, rewiring of nodes is favored, whereas for $p < 1/2$ threshold adaptations are more likely.
Notice that the model introduced in \cite{BornholRohlf00} is contained as the limiting case $p = 1$ (rewiring only and $h_i = const. = 0$ for all sites).

{\em Results.} %The core of the evolutionary algorithm is a coupling between {\em local} measurements of 
%the average dynamical activity of a site (order parameter)
%and local adjustments of either its input number or its threshold (control parameter) based on the result
%of this measurement. 
After a large number of adaptive cycles, networks self-organize into
a {\em global} evolutionary steady state.
%, which is characterized by statistical distributions
%of dynamical and topological order parameters that are stationary with respect to the evolutionary process.
An example is shown in Figure 2 for networks with $N =512$: starting from an initial value $\bar{K}_{ini} = 1$, the networks' average connectivity $\bar{K}$
first increases, and then saturates around a stationary mean value $\bar{K}_{evo}$; similar observations
are made for the average threshold $\bar{h}$ (Fig. 2, lower panel). 
The non-equilibrium nature of the system manifests itself
in limited fluctuations of both $\bar{K}$ and $\bar{h}$ around $\bar{K}_{evo}$ and $\bar{h}_{evo}$.
Regarding the dependence of $\bar{K}$ with respect to $p$, we make the interesting observation that it changes
non-monotonically. Two cases can be distinguished: when $p = 1$, $\bar{K}$ stabilizes at a very sparse mean value
$\bar{K}_{evo}$, e.g. for $N = 512$ at $\bar{K}_{evo} = 2.664 \pm 0.005$. When $p < 1$,  
the symmetry of this evolutionary steady state is broken.
Now, $\bar{K}$ converges to a much higher mean
value $\bar{K}_{evo} \approx 43.5 \pm 0.3$ (for $N=512$), however, the particular value which is finally reached is {\em independent of $p$}.
The latter observation is made rigorous from measurements of $\bar{K}_{evo}$ for different $N$ over $10^6$ evolutionary steps,
after systems have reached the steady state. While $\bar{K}_{evo}$ obviously depends on the system size $N$, curves
are very flat with respect to $p$ (Fig. 3, upper four curves); the same holds for $|\bar{h}_{evo}|$ (Fig. 3, lower four curves).
%This means that there are only two universality classes ($p=1$ and $p<1$), each with own characteristic values $\bar{K}_{evo}(N)$ 
%and $\bar{h}_{evo}(N)$. 
On the other hand, {\em convergence times} $T_{con}$ 
needed to reach the steady state are strongly influenced by $p$: $T_{con}(p)$ diverges when $p$ approaches $1$ (compare Fig. 2
for $p = 0.99$). We conclude that $p$ determines the {\em adaptive time scale}. This is also reflected by the stationary
in-degree distributions $p(k_{in})$ that vary considerably with $p$ (Fig. 4); when $p \to 1$, these distributions become very
broad. The numerical data suggest that a power law
\begin{equation}
\lim_{p \to 1} p(k_{in}) \propto k_{in}^{-\gamma}
\end{equation} 
with $\gamma \approx 3/4 \pm 0.03$ is approached in this limit (cf. Fig. 4, dashed line). At the same
time, it is interesting to notice that the evolved out-degree distributions are much narrower and completely
insensitive to $p$ (Fig. 4, data points without lines).

How can one understand the emergence of broad in-degree distributions for with increasing $p$? Evidently,
life times of both low thresholds $|h_i| \approx 0$ and high thresholds $|h_i| \gg 0$  
become long for $p \to 1$. Since sites with low thresholds
tend to be active and hence, on average, loose links, while sites with high thresholds tend to freeze
and hence, on average, aquire new links, we would indeed expect that $p(k_{in})$ is broadened for $p \to 1$.
On the other hand, for $p \to 0$, frequent adaptive changes of thresholds prevent long sequences
of both frozen or highly active states, and hence make emergence of strong local wiring heterogeneities 
less probable.
%long sequences of subsequent link additions
%at "hub nodes",
%and hence no power-law tail is present. 
If this idea is correct, we would expect
that, in the limit $p \to 1$, the in-degree of sites should exhibit a strong positive correlation to their thresholds, while for
$p \to 0$ these correlations should be less pronounced. This is indeed exactly what we observe. For $p = 0.99$,
the average in-degree $\langle k_{in}\rangle$ of a given node, as a function of its threshold $|h|$,
 shows a steep increase, while the corresponding curve is relatively flat for $p = 0.3$ (Fig. 5). 

An interesting question is whether the networks with $p < 1$ still approach a self-organized critical state for 
large $N$, as it was found for the case $p = 1$ \cite{BornholRohlf00}. Since networks now evolve 
more densely wired, non-trivial topologies,
this question has to be answered by application of a {\em dynamical} criterion. For this purpose, we studied {\em damage
spreading}: after each adaptive step, dynamics was run from an initial system state $\vec{\sigma}$ and again from a
direct neighbor state $\vec{\sigma}^{\,\prime}$ differing in one bit; after $t = 200$ updates, the Hamming distance $d$ between both
trajectories was measured and the average fraction of damaged nodes $\bar{y}(t) =: d/N$ was determined. Figure 6 shows $\bar{y}$, averaged over
$10^6$ evolutionary steps, as a function of $N$. We find that the finite networks investigated here are all supercritical,
however, $\bar{y}$ decreases monotonically with $N$. The average scaling behavior can be fit by
\begin{equation}
y(N) \approx a\cdot [\ln(N)]^{-\beta}
\end{equation}
with $a = 0.77 \pm 0.02$ and $\beta = 0.917 \pm 0.01$. This dependence indicates that $\bar{y} = 0$, i.e. the critical 
transition form chaotic to frozen dynamics,
is approached for large $N$. Notice, however, that convergence is logarithmic, whereas for 
$p =1$ power laws were found \cite{BornholRohlf00,LiuBassler2006}. Again, this indicates that $p <1$ networks constitute an entirely new universality class.

To summarize, we studied a model of network evolution that couples both rewiring of inputs
and adaptation of activation thresholds to local measurements of a dynamical order parameter. 
A control parameter $p$ determines the probability of threshold adaptations vs. link rewiring.
While for $p =1$ (rewiring only, no threshold adapttation) networks evolve a self-organized critical state with a 
sparse average connectivity $\bar{K}_{evo} \approx 2$, for any $p < 1$ (both rewiring and threshold adaptation)
networks evolve a significantly more dense wiring, with broad heterogeneous in-degree distributions approaching a power-law
$\sim k_{in}^{-3/4}$ for $p \to 1$. In this limit, time scale separation between rare threshold adaptations and frequent rewiring
leads to emergence of strong correlations between thresholds and in-degree. We presented evidence that, in the limit of large $N$,
networks logarithmically approach a self-organized critical state.

Our model presents a novel mechanism leading to co-evolution
 of topological {\em and} dynamical heterogeneity  with robust homeostatic regulation,
the latter reflected e.g. by the insensitivity of the evolved average connectivity
with respect to $p$. Since similar - seemingly contradicting -
observations are also made in experimental data of, e.g., gene regulatory networks \cite{RamoeKesseliYli06,Tong2004}, it is interesting to
speculate that similar mechanisms might be at work in the evolution of biological networks.

\end{document}